\documentclass{article}
\usepackage{tabularx}
\usepackage[dvips]{graphicx}
\usepackage{amsmath}
\usepackage{times}
\usepackage{overcite}
\begin{document}

\def\runtitle{New Window Resonance in the Potassium $3s$
Photoabsorption Spectrum}
\def\runauthor{Michi \textsc{Koide}  \textit{et al}}

\title{%
New Window Resonances in the Potassium $3s$ Photoabsorption Spectrum
}

\author{
Michi  \textsc{Koide} \and 
Fumihiro \textsc{Koike}$^{1}$ \and 
Ralf \textsc{Wehlitz}$^{2}$
     \thanks{Present address: Synchrotron Radiation Center,
     3731 Schneider Dr., Stoughton, WI-53589, U.S.A.} \and 
Ming-Tie \textsc{Huang}$^{2}$
     \thanks{Present address:Chemistry Division, Argonne National Laboratory
     9700 South Cass Ave. Argonne, Illinois 60439, U.S.A.} \and 
Tetsuo \textsc{Nagata}$^{3}$ \and 
Jon C. \textsc{Levin}$^{4}$ \and 
Stephan \textsc{Fritzsche}$^{5}$ \and 
Brett D. \textsc{DePaola}$^{6}$ \and 
Shunsuke \textsc{Ohtani} \and  
and Yoshiro \textsc{Azuma}$^{1}$\thanks{E-mail address:azuma@.post.kek.jp}
}

\maketitle
\small{Institute for Laser Science, University of
Electro-communications, Chofu, Tokyo 182-8585 \\
$^{1}$Information Networking Center, Kitasato University,
Sagamihara, Kanagawa 228-8555 \\
$^{2}$Photon Factory, Institute for Materials Structure Science,
KEK, Tsukuba, Ibaraki 305-0801 \\
$^{3}$Department of Science and Technology, Meisei University,
Hino, Tokyo 191-0042 \\
$^{4}$Department of Physics and Astronomy, University of
Tennessee, Knoxville, Tennessee 37996-1200, U.S.A. \\
$^{5}$Fachbereich Physik, Universitaet Kassel, Heinrich-Plett-Str.
40, D-34132 Kassel, Germany  \\
$^{6}$Physics Department, Kansas State University, Manhattan,
Kansas 66506, U.S.A.}


\begin{abstract}
The photoion spectrum of atomic potassium was measured over the $3s
\rightarrow np$ excitation region with the photoion time-of-flight method
and monochromatized synchrotron radiation.
An unusual spectrum with paired windows structure was found instead of a
simple regular Rydberg series.
Such subsidiary windows have not been observed in  the $3s\rightarrow np$
resonances of Ar, which has a closed outer shell.
Based on Dirac-Fock calculations,
the dual window structure at 36.7 eV and at 37.4 eV was assigned
to the $3s^{-1}3p^{6}4s4p$ resonance.
The line shape can be fitted by Fano's formula and
the Fano parameters were obtained.
\end{abstract}




\section{Introduction}
Autoionizing resonances of atoms, involving configuration mixing of discrete
excited states with direct ionization continua, have been studied
intensely~\cite{Fano61, Fano65, Shore67, Samson63,
Cooper62, Madden63, Madden69, Codling71, Ederer71, Yoshino79,
Baig86, Sorensen94, Canton00, Caldwell00, Liu99,Schulz96, Mansfield84,
vanKampen97,Lagutin99, Kjeldsen99, Wilson99}
since the first measurement of the double photoexcitation resonances of
helium with synchrotron radiation.
In particular, the subvalence $s$-shell photoexcitation in rare gas atoms
~\cite{Cooper62, Madden63, Madden69, Codling71, Ederer71, Yoshino79, Baig86,
Sorensen94, Canton00, Caldwell00,
Liu99} and closely related species
~\cite{vanKampen97,Lagutin99,Kjeldsen99,Wilson99}
have been noted for the window type resonances, appearing  as dips
in the continuum background of the photo-absorption spectrum.
Window resonances are extreme cases of the Beutler-Fano profile with
profile index $q$ close to zero.  Such window type resonance can be of
particular interest for resonances with very small oscillator-strengths
since they may nevertheless appear conspicuously in the
spectrum through strong interactions with the continua;
and, furthermore, the spectral window depth are affected
strongly by the phases of channel wavefunctions.

In the present work, we performed charge-state resolved photoion yield
spectrscopy measurements of the potassium $3s \rightarrow np$ resonances
with
monochromatized synchrotron radiation and photoion time of flight (TOF)
spectroscopy technique.  Subsequently, multi-configuration Dirac-Fock
(MCDF) calculations~\cite{Parpia96} were performed in order to understand
the structure of the spectrum.
It was found that the $3s^{1}3p^{6}4snp $ series appeared prominently as
window resonances in the direct $3p$ ionization continua.

\section{Experiment and Results}
The photoion yield spectra were measured at the 2.5 GeV electron storage
ring of the Photon Factory, KEK in Tsukuba, Japan.
The BL-3B bending magnet beamline with a 24m spherical grating
monochromator (24-m SGM~\cite{Yagishita91,Masui92}) was employed.
A 200 \textit{l}/mm laminar type grating was used for this
experiment
with 100 $\mu$m symmetric slit setting which resulted in 50 meV resolution.
Monochromatized synchrotron radiation of 35 - 45 eV irradiated atomic potassium
which were created by a resistive heating metal vapor oven and led into the
collision chamber from
the bottom. The background pressure in the experimental chamber was kept
below 1.0$\times$10$^{-7}$ Torr.
The photoions were detected by a time of flight (TOF) mass
spectrometer~\cite {Sato85}  mounted
perpendicular to both the atomic and photon
beams, operating in a pulsed extraction mode.
The charge-state resolved photoion yield spectra of K$^{+}$ and K$^{2+}$
were measured by gated collection of the photoion TOF data obtained in each step
of the photon energy scan.
The photoion counts were normalized by the
photon flux of the synchrotron radiation monitored
with the drain current from the post-focusing mirror.
The photon energy calibration was accomplished by
measuring the spectra of some rare gases~\cite{Yoshino79,Sorensen94}.

\begin{figure}[h]
     \begin{center}

\includegraphics[width=13cm,keepaspectratio,clip]{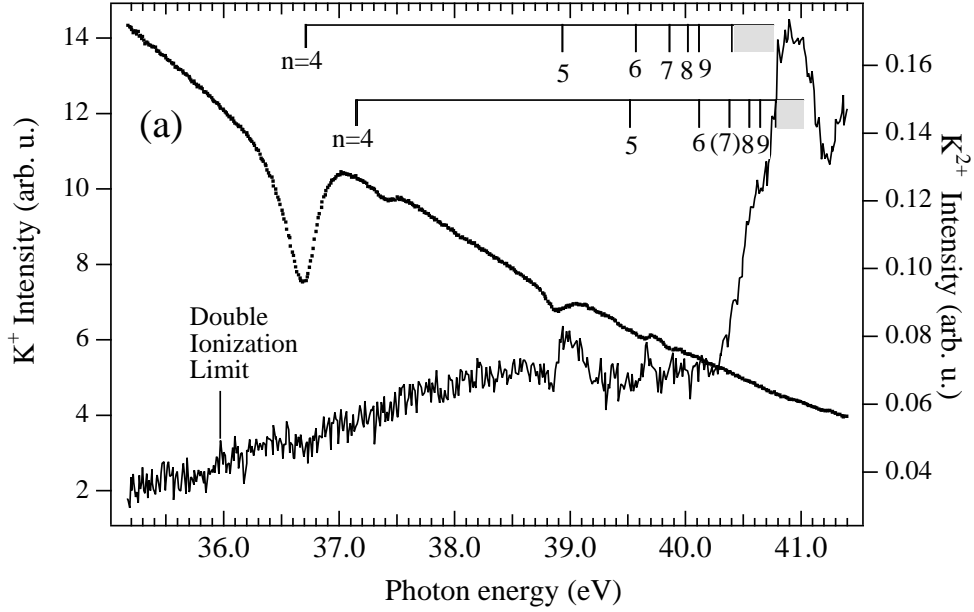}
     \end{center}
    \caption{Partial photoion yields of K$^{+}$ (dots) and K$^{2+}$ (solid
line).
             The scale for the K$^{+}$ yield is shown on the left-hand side
while
             the scale for the K$^{2+}$ yield is shown on the  right-hand
side.
             The theoretical $3s \rightarrow np$ resonance energies are
indicated
             in the upper part of the figure.}
    \label{fig:K_big}
\end{figure}

The relevant part of the charge-state resolved photoion yield spectrum is
shown in Fig.\ref{fig:K_big}.
The singly charged K$^{+}$ yield spectrum appears very similar
to the total absorption spectrum since the K$^{2+}$ yield is very small
in the photon energy region of interest.  Some window type features
attributed to K$~3s \rightarrow np$ excitations are found in the K$^{+}$
yield spectrum.
However, their appearance is in contrast to the Ar~
$3s \rightarrow np$ spectrum, which shows a very sharp and regular Rydberg
series~\cite{Baig86} up to high $n$.
The potassium resonances are broader with a less clear Rydberg
series structure, disappearing at lower $n$.
This is true not only for neutral potassium atoms but also for potassium ions
as observed previously~\cite{vanKampen97,Lagutin99,Kjeldsen99,Wilson99}.
Since K$^{+}$ is isoelectronic to Ar, it can be inferred that the presence of a valence
$4s$ electron does not play a  decisive role in this difference.
However, the most striking feature of the K$^{+}$ yield spectrum is a paired
windows structure, with the deepest and lowest feature at 36.7 eV and a
smaller
one right next to it at 37.4 eV, clearly too close to be the next member of
the Rydberg series.
Since neither the K$^{+}$ nor the Ar$~3s \rightarrow 4p$ resonances
exhibit any subsidiary dip next to the main window resonance, this can be interpreted
as due to the presence of the $4s$ electron.

The resonances in the K$^{+}$ yield spectrum were fitted with a
Fano profile for an isolated discrete state interacting with a single
continuum~\cite{Fano61,Fano65}, convoluted with
the 50 meV bandpass of the monochromator. The results are
shown in Table \ref{tab:fit}.
\begin{table}
  \caption{Fano parameters obtained using Eq.(\ref{eq:Fano1}).
                A monochromator bandpass of 50 meV was taken into account in
the fitting procedure.
                It was carried out only quartet windows without $3s
\rightarrow 4p $ window.
                Error limits were obtained from  the standard deviation.
          }

 \begin{tabular}{@{\hspace{\tabcolsep}\extracolsep{\fill}}cccccccc}
\hline
     $n$             &  4  &  4\footnote{$^{2}\!P$} &   5  &  6  &  7  &  8
& 9    \\
\hline
  $E_{r}$ (eV)         &36.72(5)&37.4(6)  &38.90(8)
&39.6(1)&39.91(8)&40.0(2)&40.1(1) \\
  $\Gamma$(eV)     &0.21(5)  &0.15(9)  &0.16(8)   &0.09(9)  &0.07(9)
&0.06(9)  & -            \\
  $\Gamma$\footnote{Present calculation} &0.157&0.136 &   -
&    -&    -&   - &  -    \\
  $q$
&0.23$\pm$0.01&0.6$\pm$0.1&0.14$\pm$0.03&1.0$\pm$0.2&0.8$\pm$0.1&0.8$\pm$0.1
&0.6$\pm$0.3 \\
\hline
  \end{tabular}
  \label{tab:fit}
\end{table}
The photoion yield, $\sigma(E)$, is given by
  \begin{equation}
      \sigma(E)=\sigma_{a} \frac{(q+ \epsilon)^{2}}
                                        {\epsilon^{2}+1}
                    + \sigma_{b}
                             \label{eq:Fano1}.
\end{equation}
Here, $\epsilon(E)$ is the reduced energy and is given
by $\epsilon(E) =(E-E_{r})/\frac{1}{2} \Gamma $,
where $E$ is the incident photon energy, $E_r$ is the resonance
energy and $\Gamma$ is the FWHM (full width at half maximum).
$\sigma_{a}$ is the depth of the minimum
associated with the resonance, and represents the portion of the continua
that interacts with the discrete state.
$\sigma_{b}$ is the portion of the continua that does not
interact with the discrete state.  The profile index
{\it q} gives the shape of the resonance. When {\it q} is smaller than 1,
the resonances form a dip like structure, i.e. a window resonance.

\section{Calculations and Results}
A series of MCDF calculations were performed by the
programs GRASP92~\cite{Parpia96} and RATIP~\cite{Fritzsche01}.  
We used also an older version of GRASP code (GRASP2)
for the non-relativistic symmetry indices such as the
total spin $S$ and the total orbital angular momentum $L$,
which are useful to refer the configuration state functions
expressed in terms of non-relativistic scheme of conventions.
The total electronic energies of the K ground state,
two series of $3s \rightarrow np$ excited  states that could be
assigned as either quartet $P$ or doublet $P$ in the non-relativistic sense
were calculated.
And the potassium ionic states with a $3s$ or $3p$ hole were also
calculated.
Differences in the energy of the ground state and the excited
state energies correspond to the resonance energies.  Also, the oscillator
strengths of $3s \rightarrow np$ excitations as well as the Auger rate of
$3s \rightarrow np$ excited states were obtained.
Relevant results of the calculations are shown in Table {\ref{tab:cal}};
we found two series of photo-excitations that are comparable
with the present experiment.
Although the total spin multiplicities are not good indices in a 
relativistic regime, we may still assign the states in terms of the
total spin multiplicities as of the leading configurations in
light atoms.
\begin{table}[htl]
  \caption{The calculated and the observed energies (in eV) of the
$3s^{1}3p^{6}4snp$ series  along with the oscillator strengths of the lower
     $J$ states. The observed energies of higher $^{2}\!P$
     and $3s$ direct ionization threshold are obtained from the K$^{2+}$
yield spectrum.
     }
  \begin{tabular}{@{\hspace{\tabcolsep}\extracolsep{\fill}}ccccccc}
\hline
main configuration&\multicolumn{3}{c}{Lower Series~$^{4}\!P$(93\%)$+^{2}\!P$(7\%)}&
\multicolumn{3}{c}{Higher Series~$^{2}\!P$(93\%)$+^{4}\!P$(7\%)} \\ \cline{2-4}\cline{5-7}
       ~~                  &   obs. &    cal.   &{\it gf}&  obs.   &  cal.
&{\it gf}\\
\hline
$3s^{1}3p^{6}4s4p$        &  36.72(5)  &  36.67  &$2.5\times10^{-5}$&
37.4(6)  & 37.11  &$2.1\times10^{-2}$\\
$3s^{1}3p^{6}4s5p$        &  38.90(8)  &  38.89  &$4.1\times10^{-6}$&
38.9(7)  & 39.04  &$3.6\times10^{-3}$\\
$3s^{1}3p^{6}4s6p$        &  39.6(1)~~  &  39.53  &$1.6\times10^{-6}$&
39.7(7)  &39.59  &$1.3\times10^{-3}$\\
$3s^{1}3p^{6}4s7p$        &  39.91(8)  &  39.82  &$8.4\times10^{-7}$&
39.9(9)  &  -       &$6.8\times10^{-4}$\\
$3s^{1}3p^{6}4s8p$        &  40.0(2)~~  &  39.98  &$4.7\times10^{-7}$&    -
&39.00  &$3.9\times10^{-4}$\\
$3s^{1}3p^{6}4s9p$        &  ?~40.1(1)~~  &  40.08  &$3.1\times10^{-7}$&    -
&40.09  &$2.5\times10^{-4}$\\
$3s^{1}3p^{6}4s\epsilon{p}$&?~40.28(5)&40.36&   -      & ?~40.67(5)   &  40.74
& - \\
\hline

  \end{tabular}
  \label{tab:cal}
\end{table}

The term purity in $LS$-coupling for these states is
approximately 93 \%  based on the our  calculation results;
i. e., the major contributions are from the quartet for the lower
series in Table \ref{tab:cal} and from the doublet for the
higher series in Table \ref{tab:cal}.
And further on, this means that 7 \% of the configuration
state component is of doublet in the lower series,
and, consequently, that, in this series, the photo-excitation from the ground states is 
realized though this minor component.
We would call, in the Table \ref{tab:cal}, the lower series 
as quartet and the higher series as doublet, here after, 
for the sake of convenience.
The calculated oscillator strengths are quite small in both 
doublet and quartet $3s \rightarrow 4p$ resonances; 
we obtained $2.1\times10^{-2}$ for
$^{2}\!P_{1\!/\!2}$ and $2.5\times10^{-5}$ for $^{4}\!P_{1\!/\!2}$
(see also Table \ref{tab:cal}).
Due to these very small oscillator strengths, the $^{4}\!P$
and $^{2}\!P$ resonances in the photoionization spectra may appear as
windows with the reduction of photoabsorption over the resonance positions.
However, the window depths are not of the direct concern
to the values of the oscillator strengths; the depths are influenced 
strongly by the phases of ionization channels. 
Nevertheless, the larger $q$ value for the $^{2}\!P$ than for the
$^{4}\!P$ may reflect the small yet greater oscillator strength of the
doublet.

Significant  $3s^{2}3p^{4}3d^{1}4snp$ configuration mixing
with the $3s$-hole states $3s^{1}3p^{6}4snp$
were found.
The magnitude of mixing reaches almost 25 {\%} of the total
occupation, causing the total state-energy to decrease
by almost 6.47 eV.
Also, a significant contribution from the $3d\,^{2}$
configurations were obtained by incorporating the $d$-orbitals up to $4d$ in
the MCDF wave function expansions.
The $3p$ or $4s$ ionization energies can be used as benchmarks to
assess the accuracy of the present calculation scheme.
We have calculated the total  energies of
K$^{+} ~3s^{2}3p^{5}4s ~^{3,1}\!P^{o}$, K$^{+}
~3s^{2}3p^{6}$, and K$^{2+} ~3s^{2}3p^{5}$
from which the ionization threshold energies
with regard to the K ground state are obtained.
As seen from Table \ref{level}, the results are in good agreement with previous
data~\cite{Sugar85,Catalan58a}
and support the reliability of the present calculation.
\begin{table}
\caption{Our MCDF calculated results for the ionization thresholds are shown
              in eV unit.
              }
\label{level}
  \begin{tabular}{@{\hspace{\tabcolsep}\extracolsep{\fill}}lccccccc}
\hline
 &\multicolumn{7}{c}{configuration} \\ \cline{2-8}
&$3p^{6}$&$3p^{5}4s ~^{3}\!P_{2}$&$3p^{5}4s ~^{3}\!P_{1}$&$3p^{5}4s
~^{3}\!P_{0}$&$3p^{5}4s ~^{1}\!P_{1}$&$3p^{5} ~^{2}\!P_{3\!/\!2}$&$3p^{5}
~^{2}\!P_{1\!/\!2}$\\
\hline present&4.18&24.63&24.75&24.89&25.11&35.53&35.81\\
previous*&4.341&24.49&24.58&24.82&24.98&35.97&36.24\\
\hline
\end{tabular}
\medskip
*Sugar and Corliss~\cite{Sugar85} and Catal\'{a}n and Rico~\cite{Catalan58a}
\end{table}

\section{Discussion}
Based on our calculated photon energy positions, the
deeper window can be assigned as due to the
$3s^{1}3p^{6}4s4p$ and most of the series of windows above the lowest dual
windows
structure can be attributed to $3s^{1}3p^{6}4snp$ resonances.
The first excitation state cause  the lowest dual window.
Comparing the observations with the calculations, the dual window may be
attributed to the resonances $3s \rightarrow 4p$ $^{4}\!P$ and $^{2}\!P$.
The energy difference of 0.7 eV for the neighbored windows is important for this
assignment.
Furthermore,  the resonance width calculated by RATIP
is close to the observation (See Table\ref{tab:fit}).
The calculation predicted another $3s^{1}3p^{6}4s4p$ 
doublet state by 2.38 eV above the deepest window,
which was invisible in the present experimental spectra.
This doublet $P$ state and another doublet $P$ state which corresponds to 
the first entry of the "higher series" form a pair of doublet states with
the same configuration $3s^{1}3p^{6}4s4p$.
The energy difference of these doublet $P$ states 
 2 eV is considered as mainly from the term splitting between
$4s4p$ singlet and triplet energies. 

The main decay channel from the $3s \rightarrow np$ excited
states is the following:\\
~~~~~$ 3s^{1}3p^{6}4snp  \longrightarrow
                    3s^{2}3p^{5}4s +  e^{-}$\\
in this photon energy region according to our Auger rate calculations using
RATIP~\cite{Fritzsche01}.
On the other hand, for the continuum contributions, the predominant channel
is the $3p$ direct ionization:\\
~~~~~$3s^{2}3p^{6}4s  +  h\nu  \longrightarrow
                         3s^{2}3p^{5}4s  + e^{-} $\\
Since the final state in the predominant decay channel and the predominant
direct ionization channel have the same configuration, they may interfere
and mix strongly if their symmetries match.
Therefore both the quartet and doublet terms in the
intermediate K$~  3s^{1}3p^{6}4snp$ state
configurations require components in the continua that satisfy the symmetry
condition, in order to form autoionizing resonances.
Since the final ionic states have open shells and have two possible
$j$ for a $3p$ electron, $j=1/2$ and $3/2$,
the $LS$ term of final state cannot be decided uniquely.

Two thresholds of $3s$ direct ionization are clearly observed
in the K$^{2+}$ yield spectrum at 40.28 eV
($3s^{1}3p^{6}4s~ ^{3}\!S$) and at 40.67 eV ($3s^{1}3p^{6}4s~ ^{1}\!S$).
The energy positions are in good agreement with
the calculations (See Table\ref{tab:cal}).
These correspond to the series limit $3s \rightarrow np$,
$3s^{1}3p^{6}4snp~ ^{4}\!P$ and $3s^{1}3p^{6}4snp~ ^{2}\!P$.
The limits did not appear in the K$^{+}$ spectrum.
There is no clear feature at the double ionization threshold
35.971 eV~\cite{Sugar85,Catalan58a} in the K$^{2+}$ spectrum.
Below the $3s$ direct ionization threshold, a peak-like Rydberg series was
found.
This series may be attributed to the doublet term of $3s \rightarrow np$
$(n\ge 5)$ excitation.
Even though direct double photoionization is energetically
possible above 35.971 eV~\cite{Sugar85,Catalan58a},
it is very weak compared to the single ionization process in this energy
region.
Hence, the $^{2}\!P$ series, with a much larger oscillator strength,
appear as a peak like Rydberg series in the K$^{2+}$ yield spectrum.

\section{Summary}
In summary, the $3s \rightarrow np$ photoexcitation resonances of atomic
potassium were observed between 35.5 eV and 40 eV incident photon energy
through charge state resolved photoion spectroscopy.
New paired windows  and series of windows were observed for the first time and
they were assigned as $3s^{1}3p^{6}4snp$.
The assignments have been made based on
the results of our MCDF calculations.
The valence $4s$ electron of K induces the term energy
splitting by electron correlation, and hence, the two states
could be resolved.
This new dual window feature could be attributed to $3s^{1}3p^{6}4s4p$
$^{4}\!P$ and $^{2}\!P$.
Since the $^{4}\!P$ assignment would mean an $LS$ forbidden transition,
it requires further scrutiny, although the good agreement between the calculated
and observed energy positions
is very suggestive.  More thorough investigations both theoretically and
experimentally are needed for an unambiguous interpretation of this process.

Support to R.W. from the
Japan Society for the Promotion of Science is
gratefully acknowledged.  This
work was performed under the approval of the
Photon Factory Program Advisory
Committee (Proposal No.\ 98G028).
Discussions with X.-M.\ Tong, D.\ Kato,
Y.\ Suzuki, C.\ Yamada, and T.\ Watanabe  are
appreciated.

\end{document}